     \documentclass[prb,preprint,showpacs,preprintnumbers,amsmath,amssymb]{revtex4}

\usepackage{graphicx}
\usepackage{dcolumn}
\usepackage{bm}

\def\vq{{\bf q}}

\def\vk{{\bf k}}
\def\vQ{{\bf Q}}

\newcommand{\eq}[1]{Eq.~(\ref{#1})}
\newcommand{\fig}[1]{Fig.~\ref{#1}}
\newcommand{\be}{\begin{equation}}
\newcommand{\ee}{\end{equation}}
\newcommand{\bea}{\begin{eqnarray}}

\newcommand{\eea}{\end{eqnarray}}
\newcommand{\bean}{\begin{eqnarray*}}
\newcommand{\eean}{\end{eqnarray*}}
\newcommand{\bfi}{\begin{figure}}
\newcommand{\efi}{\end{figure}}
\newcommand{\bc}{\begin{center}}
\newcommand{\ec}{\end{center}}
\newcommand{\ba}{\begin{array}}
\newcommand{\ea}{\end{array}}
\def\bra{\langle}
\def\ket{\rangle}

\begin{document}


\title{Multilayer cuprate superconductors as possible systems 
described by resonating-valence-bond and antiferromagnetic orders
}

\author{Hiroyuki Yamase$^1$}
\author{Masanao Yoneya$^2$}
\author{Kazuhiro Kuboki$^2$}
\affiliation{$^1$ National Institute for Materials Science, Tsukuba 305-0047, Japan\\
$^2$Department of Physics, Kobe University, Kobe 657-8501, Japan}

\date{\today}

\begin{abstract}
Coexistence of antiferromagnetism 
and $d$-wave superconductivity within a CuO$_{2}$ plane 
was recently observed in a wide doping region for multilayer high-temperature cuprate 
superconductors.  
We find that the experimental phase diagram is well 
reproduced in the slave-boson mean-field scheme of the two-dimensional $t$-$J$ model 
by including antiferromagnetic order.  
We argue that weak three dimensionality coming from 
a multilayer structure is sufficient to stabilize antiferromagnetic order and its 
coexistence with superconductivity.  
\end{abstract}

\maketitle

\section{Introduction} 
High-temperature superconductivity (SC)  is one of the 
most fascinating topics in physics. 
The discovery of the magnesium diboride in 2001 (Ref.~\onlinecite{nagamatsu01})  
immediately generated very intensive studies. 
The recent discovery of ferropnictides\cite{kamihara08} also attracts 
tremendous attention, opening a new branch in 
condensed matter physics.
It is however still  only cuprate superconductors especially multilayer 
cuprates such as 
HgBa$_{2}$Ca$_{n-1}$Cu$_{n}$O$_{y}$ and TlBa$_{2}$Ca$_{n-1}$Cu$_{n}$O$_{y}$ 
with $n \geq 3$ 
that achieve superconducting transition temperature ($T_{c}$) 
more than 100 K (Ref.~\onlinecite{iyo07});  we mean 
by "multilayer" three or more layers in a unit cell in this paper. 
For such multilayer cuprates, 
the previous theoretical studies\cite{chakravarty04,zaleski05,mori05} 
payed attention to the description of a superconducting phase coherence 
between the layers, including a possible charge imbalance between the layers in a unit cell. 
In those studies there was a tacit assumption that the property of each CuO$_{2}$ 
plane is essentially the same as that in single- and bi-layer cuprates 
in the sense that AF is realized for very low doping and is replaced by 
SC for moderate doping without the coexistence with AF within a CuO$_2$ plane. 

However, the recent NMR measurements for multilayer cuprate 
superconductors\cite{mukuda08,shimizu09,shimizu09a} revealed 
the phenomena very different from those in single- and bi-layer cuprates:  
antiferromagnetism (AF) in the Mott insulator survives up to rather high carrier doping 
and coexists with a superconducting state. 
The coexistence was  due to not a proximity effect between the layers, but 
a phase transition within a CuO$_{2}$ plane at low temperatures. 

This observation sharply contrasts with a widely accepted viewpoint that 
AF in the Mott insulator is rapidly suppressed  by a tiny amount of carrier doping
and the $d$-wave superconducting state is instead stabilized in a wide doping region, typically in $5-30 \%$. 
The latter viewpoint is based on the data for single- and bi-layer cuprates 
such as La- and Y-based compounds, for which a huge number of studies 
have been performed last about 25 years in a systematic way. 
Compared with those materials, multilayer cuprates are much less investigated. 
They contain completely flat CuO$_{2}$ planes 
with a perfect square lattice and are known to be free from disorder, 
in contrast to La- and Y-based cuprates. 
They also achieve higher $T_{c}$ than that for single- and bi-layer cuprates.\cite{iyo07} 
In this sense multilayer cuprates can be viewed as an ideal system to study the 
mechanism of high $T_c$. 
Nonetheless, the basic theoretical framework for multilayer 
cuprates has not been identified so far.

Moreover the origin of the pseudogap\cite{timusk99} is still a central issue on 
high-temperature SC and a clue to resolve it is highly desired. 
Therefore it is of great importance to  shed light on an issue  
of the pseudogap in multilayer cuprates from a theoretical point of view, 
which may in turn provide a crucial insight into the pseudogap in 
single- and bi-layer cuprates. 

In this paper, 
we explore the basic theoretical framework which captures 
the essential features recently reported for the 
multilayer cuprates.\cite{mukuda08,shimizu09,shimizu09a}  
We invoke Anderson's resonating-valence-bond (RVB) scenario,\cite{anderson87} 
that is, the undoped CuO$_{2}$ plane is assumed to be in the RVB spin liquid state 
and the preexisting spin singlet pairs can become charged superconducting pairs 
once they are mobile by carrier doping. 
This idea was well described in terms of the two-dimensional (2D) $t$-$J$ model. 
In particular, the slave-boson mean-field theory\cite{kotliar88,suzumura88}  
and the gauge theory,\cite{nagaosa90} which takes low-energy fluctuations around the mean-fields into account, turned out to capture many important properties 
of single- and bi-layer cuprate superconductors.\cite{lee06,yamase0607,ogata08}  

In the standard RVB framework, AF is assumed to be strongly fluctuating, 
not to be ordered. 
However, antiferromagnetic order can be easily stabilized 
in the presence of weak three dimensionality coming from a multilayer structure. 
This effect can be incorporated phenomenologically 
by including AF as a possible mean-field order parameter within a single-layer model. 
Such a calculation was performed in the slave-boson 
mean-field scheme\cite{inaba96,yamase04a}  
and variational Monte Carlo\cite{himeda99,shih04} for the 2D $t$-$J$ model 
in a different context, showing that AF extends to a high doping rate and coexists with SC. 
However, the coexistence obtained previously was found to be substantially suppressed 
by the presence of a long-range hopping amplitude, 
leading to a pure antiferromagnetic state.\cite{yamase04a,shih04} 
Moreover, the previous theoretical results\cite{inaba96,yamase04a} 
typically exhibited reentrant behavior of 
the critical temperature of AF at low temperatures, which  
was not observed experimentally.\cite{mukuda08}  

We perform the slave-boson mean-field analysis of the $t$-$J$ model 
by including antiferromagnetic order. 
While we analyze the 2D model and do not take multilayer degrees of freedom 
into account, we consider that a special feature of multilayer cuprates 
is included phenomenologically by allowing AF order as a possible mean field 
in our analysis. In this framework, important features of the phase diagram for 
multilayer cuprates are obtained.\cite{kuboki10}   
We argue that the essential difference between multilayer cuprates and 
single- and bi-layer cuprates lies in the presence of weak three dimensionality 
and thus the phase diagram of the former is well described by both RVB and AF, 
while that of the latter is simply 
by the RVB as already investigated. 
In contrast to single- and bi-layer cuprates, 
the pseudogap may be substantially diminished 
in the heavily underdoped region of multilayer cuprates.

\section{Model and formalism} 
We employ the 2D $t$-$J$ model on a square lattice 
\be
{\cal H} = - \sum_{i,j,\sigma} t_{ij} {\tilde c}_{i\sigma}^\dagger {\tilde c}_{j\sigma}
+ J \sum_{\langle i,j \rangle} {\bf S}_i \cdot {\bf S}_j \,, 
\label{t-Jmodel}
\ee
where the transfer integrals $t_{ij}$ are finite for the first- ($t$), 
second- ($t'$), and third-nearest neighbor bonds ($t''$), and vanish otherwise. 
$J (>0)$ is the antiferromagnetic superexchange interaction
and $\langle i,j \rangle$ denotes the nearest neighbor bonds. 
${\tilde c}_{i\sigma}$ is the electron operator in the Fock space without 
double occupancy and we treat this condition using the slave-boson method\cite{zou88}   
by writing ${\tilde c}_{i\sigma}=b_i^\dagger f_{i\sigma}$ under 
the local constraint $\sum_{\sigma}f_{i\,\sigma}^{\dagger}f_{i\,\sigma} + b_{i}^{\dagger}b_{i}=1$ 
at every $i$ site. Here $f_{i\sigma}$ ($b_i$) is a fermion (boson) operator  that carries spin $\sigma$ 
(charge $e$); the fermions (bosons) are frequently referred to as spinons (holons). 
The spin operator  is expressed as 
${\bf S}_i = \frac{1}{2}\sum_{\alpha\beta} 
f^\dagger_{i\alpha}{\boldsymbol \sigma}_{\alpha\beta}
f_{i\beta}$ with ${\boldsymbol \sigma}$ being the Pauli matrices.

Hamiltonian (\ref{t-Jmodel}) is decoupled  by introducing the following 
order parameters:\cite{inaba96,yamase04a}  
the staggered magnetization $m = \frac{1}{2}\langle f^\dagger_{i\uparrow}f_{i\uparrow}
- f^\dagger_{i\downarrow}f_{i\downarrow} \rangle e^{i{\bf Q}\cdot{\bf r}_i}$ 
with ${\bf Q} \equiv (\pi,\pi)$; the bond order parameter for spinons and holons, 
$\langle \sum_\sigma f^\dagger_{i\sigma}f_{j\sigma} \rangle$, 
$\langle b^\dagger_ib_j\rangle$; 
we denote 
$\chi=\langle \sum_\sigma f^\dagger_{i\sigma}f_{j\sigma} \rangle$ for 
the nearest neighbor bond; 
the singlet RVB paring $\Delta_\tau = \langle f_{i\uparrow}f_{i+\tau \downarrow} 
-f_{i\downarrow}f_{i+\tau \uparrow}\rangle$ with $\tau=x, y$. 
Here we assume that all these expectation values 
are real and independent of $i$. 
We can show that the $d_{x^2-y^2}$-wave pairing state 
is the most stable, i.e., $\Delta_x = -\Delta_y \equiv \Delta_0$.  
Although the bosons are not condensed in the present mean-field scheme 
at finite temperature ($T$), 
they are almost condensed at low $T$ and 
for finite carrier doping $\delta (\gtrsim 0.02)$.\cite{inaba96}  
Hence we approximate $\langle b \rangle \approx {\sqrt \delta}$ and 
$\langle b^\dagger_ib_j\rangle \approx  \delta$. 
In principle, the so-called $\pi$-triplet pairing state can emerge for a state with 
$m\neq 0$ and $\Delta_{0}\neq 0$,\cite{psaltakis83} but turns out not to be stable 
in our model.  
Hence the free energy per lattice site is computed as 
\be
F=-\frac{2T}{N} {\sum_{\vk}}^{'} \left[ \log \left(2 \cosh \frac{\lambda_{\vk}^{+}}{2T}\right) +
\log \left(2 \cosh \frac{\lambda_{\vk}^{-}}{2T}\right) \right] +
\frac{3J}{4} \left(\chi^{2} + \Delta_{0}^{2}\right) + 2J m^{2} -\mu \delta \,,
\label{free-energy}
\ee
where $\lambda_\vk^\pm = \sqrt{(\eta_\vk^{\pm})^2+\Delta_\vk^2}$ is the spinon's 
band dispersion in the presence of $m$ and $\Delta_{0}$; 
$\eta_\vk^\pm = \xi_\vk^+ \pm D_\vk$, $D_\vk = \sqrt{(\xi_\vk^-)^2+(2Jm)^2}$, 
$\xi_\vk^\pm = (\xi_\vk \pm \xi_{\vk+\vQ})/2$, 
$\xi_\vk=-2(t\delta+\frac{3}{8}J\chi)(\cos k_x+\cos k_y)
-4t'\delta\cos k_x\cos k_y -2t''\delta(\cos 2k_x+\cos 2k_y)$,  
and $\Delta_\vk=-\frac{3}{4}J\Delta_0(\cos k_x-\cos k_y)$; 
$\mu$ and $N$ denote the chemical potential and 
the total number of lattice sites, respectively; 
the sum of momentum is taken over the magnetic Brillouin zone 
$|k_x|+|k_y| \leq \pi$. 

Because we relax the local constraint to a global one 
$\bra \sum_{\sigma} f_{i\,\sigma}^{\dagger} f_{i\,\sigma} \ket =1-\delta$ 
and $\bra b_{i}^{\dagger} b_{i} \ket =\delta$ in the present mean-field theory, 
our approximation may be reliable as long as electrons are in 
coherent motion. In the present case, since $\chi$ tends to saturate  
for $\delta \gtrsim 0.05$ [\fig{phase}(b)], 
we expect that our approximation is sufficiently reliable in such a region, 
where most of experimental data have been obtained so far.

To examine a possibility of the incommensurate antiferromagnetic instability, 
we also compute the longitudinal 
magnetic susceptibility $\chi(\vq)$ in the random phase approximation: 
$\chi(\vq)^{-1} = \chi_{0}(\vq)^{-1}  + 2J (\cos q_{x} + \cos q_{y})$ and 
\bea
&&\hspace{-5mm}\chi_{0}(\vq)=\frac{1}{4 N}
\sum_{\vk}\left[C^{+}_{\vk,\,\vk+\vq} 
 \frac{\tanh \frac{E_{\vk}}{2 T}
   -\tanh \frac{E_{\vk +\vq}}{2 T}}
   {E_{\vk}-E_{\vk+\vq}}\right. 
   +\left.C^{-}_{\vk,\,\vk+\vq}\frac{\tanh \frac{E_{\vk}}{2T}
  +\tanh \frac{E_{\vk +\vq}}{2T}}
{E_{\vk}+E_{\vk+\vq}}\right],\\
&&C^{\pm}_{\vk,\,\vk+\vq}=\frac{1}{2}
\left(1 \pm \frac{\xi_{\vk}\xi_{\vk+\vq}
 +\Delta_{\vk}\Delta_{\vk+\vq}}{E_{\vk}E_{\vk+\vq}}\right)\, .
\eea
Here $E_{\vk}=\sqrt{\xi_{\vk}^{2}+\Delta_{\vk}^{2}}$ is the spinon's   
band dispersion for $m = 0$ and 
the sum of $\vk$ is taken over the region $|k_x|, |k_y| \leq \pi$. 

The material dependence of cuprate superconductors can be taken 
into account mainly by different choices of $t'$ and $t''$.\cite{tanamoto93,feiner96,tohyama00,pavarini01} 
Hence it is naturally expected that we could invoke specific values of $t'$ and $t''$ 
appropriate for multilayer cuprates. This could be achieved for a 
realistic multilayer model, which also contains other parameters 
such as interlayer hopping integrals, interlayer exchange interactions, 
and site potential yielding a charge imbalance between the layers in a unit cell. 
Given that there is much ambiguity about those parameters and 
that a special feature of multilayer cuprates 
is considered phenomenologically in the present analysis 
by invoking AF order as a possible mean field, 
we consider the values of $t'$ and $t''$ simply as phenomenological 
parameters to reproduce various types of the phase diagram. 

\section{Results}
\begin{figure}[ht]
\includegraphics[width=10.0cm]{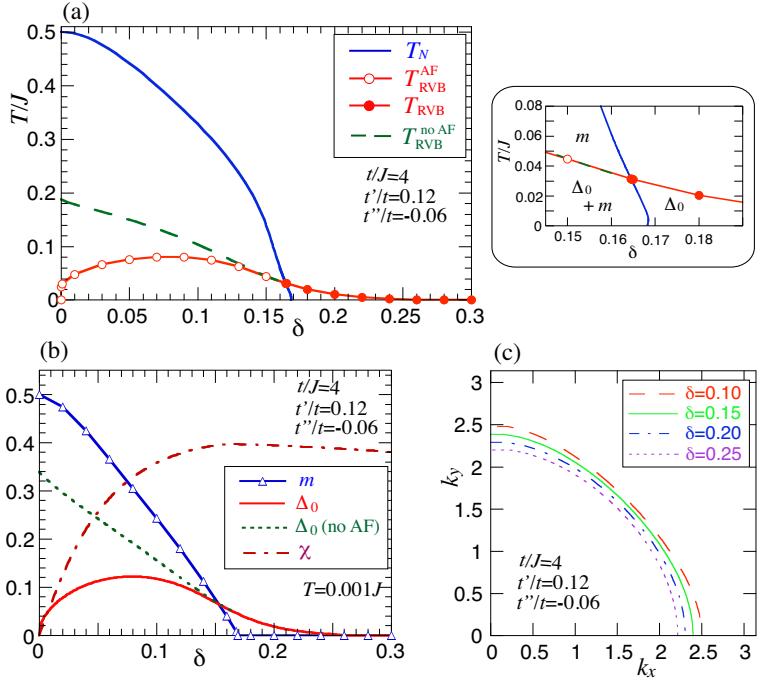}
\caption{(Color online) (a) Phase diagram in the plane of $\delta$ and $T$ 
for $t/J=4$, $t'/t=0.12$ and $t''/t=-0.06$. $T_{N}$ is the onset temperature of AF, 
and $T^{\rm AF}_{\rm RVB}$ ($T_{\rm RVB}$ and $T^{\rm no\, AF}_{\rm RVB}$) 
is that of singlet pairing in the presence (absence) of AF. 
The right-hand panel magnifies the region around the tetracritical point; 
the antiferromagnetic and superconducting states are denoted by 
$\lq\lq m"$ and $\lq\lq \Delta_{0}"$, respectively, and their coexistence is by 
$\lq\lq \Delta_{0}+m"$.   
(b) $\delta$ dependence of the order parameters at $T=0.001J$. 
(c) Fermi surfaces in the normal state for several choices of $\delta$.} 
\label{phase}
\end{figure}

We determine the mean fields by minimizing the free energy \eq{free-energy} and 
obtain the phase diagram in the plane of $\delta$ and $T$. 
The result for $t/J=4$, $t'/t=0.12$, and $t''/t=-0.06$, for 
which an electron-like Fermi surface is realized in a normal state [\fig{phase}(c)], 
is shown in \fig{phase}(a). 
The temperature $T_{N}$ denotes 
the onset of AF, and $T_{\rm RVB}^{\rm AF}$ ($T_{\rm RVB}$ and $T_{\rm RVB}^{\rm no\, AF}$)  
the onset of singlet paring in the presence (absence) of AF. 
At $\delta=0$, AF is realized  and no singlet pairing coexists. 
With carrier doping, the critical temperature of AF gradually decreases and survives 
in a wide doping region up to $\delta_{N} \approx 0.17$; 
the $\delta_{N}$ is the critical doping rate of AF at $T=0$.  
AF suppresses the formation of singlet pairing 
($T_{\rm RVB}^{\rm AF} < T_{\rm RVB}^{\rm no\, AF}$), 
but the coexistence is realized at low temperatures. 
Singlet pairing extends over a wider doping region than AF, and a pure 
$d$-wave singlet pairing state is realized at high doping rates. 
The right-hand panel in \fig{phase}(a) magnifies the region around the tetracritical point 
$(\delta_{\rm tet}, T_{\rm tet}) \approx (0.165, 0.031J)$, where the four states, 
AF, SC,  their coexistence, and the normal states, become identical. 
The order parameters $m$, $\Delta_{0}$, and $\chi$ are shown in 
\fig{phase}(b) as a function of $\delta$ at low $T$. 
Although $\Delta_{0}$ still increases with 
decreasing $\delta$ in a certain region in $\delta \lesssim \delta_{N}$, 
we see that $\Delta_{0}$ is typically suppressed by the presence of AF 
compared with $\Delta_{0}$ for $m=0$. 
Similarly the magnitude of $\chi$ is suppressed by the presence of $m$ 
and becomes zero at $\delta=0$. 
The system becomes an antiferromagnetic insulator at $\delta=0$. 

A crucial feature of \fig{phase} (a) (see the top right-hand panel) 
is that in spite of competition of AF and 
singlet pairing, AF extends to a higher doping region 
beyond the tetracritical point in the singlet pairing state. 
This is the crucial difference from the previous work\cite{inaba96,yamase04a} 
where $T_{N}$  exhibits reentrant behavior at low $T$, leading to 
$\delta_{N} <\delta_{\rm tet}$. 
The difference comes from our careful choice of the band parameters 
under the constraint $t''=-t'/2$ (Ref.~\onlinecite{andersen95}) 
such that the shape of the Fermi surface around $\delta=\delta_{\rm tet}$ 
fulfills the nesting condition of $\vq=\vQ$  
close to the nodal region of the $d$-wave pairing gap. 
Hence, the resulting static spin susceptibility $\chi(\vq)$ shows a maximum 
at $\vq=\vQ$ and is not suppressed by the onset of singlet pairing, 
leading to $\delta_{N} > \delta_{\rm tet}$.

\begin{figure} [t]
\includegraphics[width=10.0cm,clip]{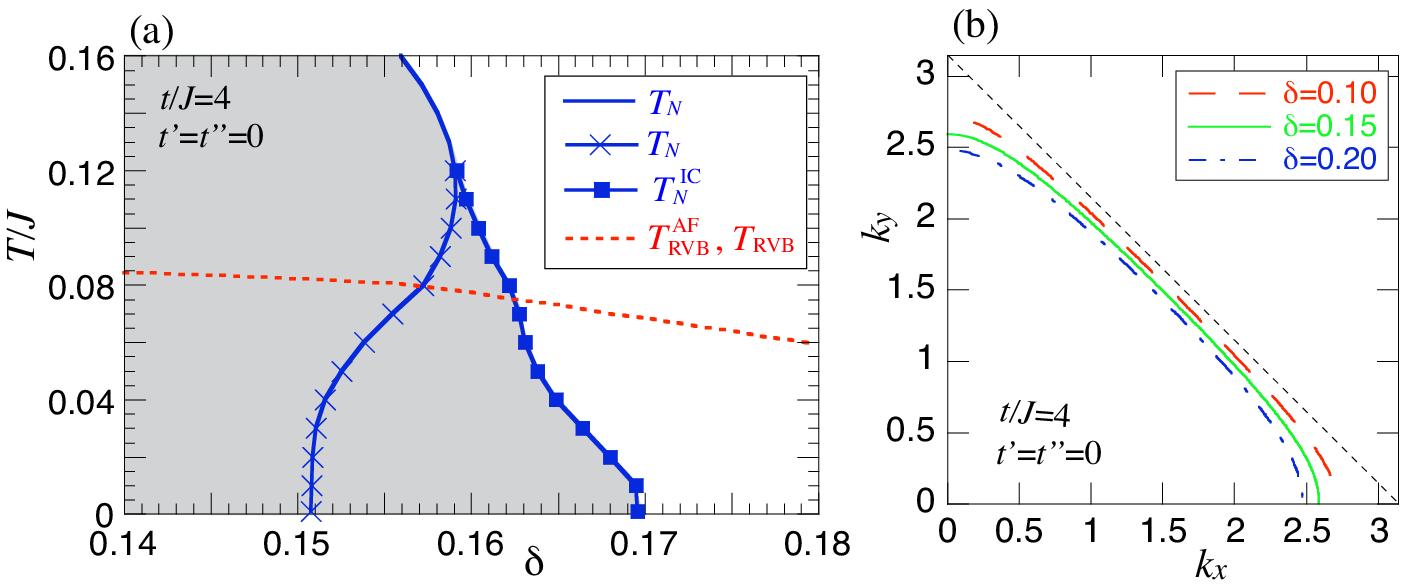}
\caption{(Color online) (a) Phase diagram in the plane of $\delta$ and $T$ for $t/J=4$, $t'=t''=0$. 
 $T_{N}^{\rm IC}$ is the onset temperature of incommensurate AF; 
other notations follow those in \fig{phase}(a). AF is realized in the shaded region. 
(b) Fermi surfaces. The boundary of the magnetic Brillouin zone is denoted by a dotted line. 
} 
\label{phase2}
\end{figure}

The careful choice of the band parameters is not a unique way to obtain a phase 
diagram similar to \fig{phase}(a) when we allow 
other magnetic orders with $\vq \neq \vQ$. 
In \fig{phase2}(a), we show a representative result. 
We first minimize the free energy \eq{free-energy} 
with respect to the mean fields $\chi$, $m$, and $\Delta_{0}$. 
We then obtain the lines of $T_{N}$, $T_{\rm RVB}^{\rm AF}$, and $T_{\rm RVB}$. 
The $T_{N}$ exhibits reentrant behavior around $\delta=\delta_{N}\approx 0.15$ 
at low temperatures, in contrast to \fig{phase}(a).  
To investigate a possibility of other magnetic orders, we check a wave vector 
at which $\chi(\vq)$ diverges. 
It turns out that a part of the line $T_{N}$ [solid line with crosses in \fig{phase2}(a)] is preempted 
by the onset of incommensurate antiferromagnetic order $T_{N}^{\rm IC}$, 
where $\chi(\vq)$ diverges at $\vq\neq \vQ$. 
The ordering wave vector $\vq$ depends strongly on $\delta$ and $T$. 
The most crucial point in \fig{phase2}(a) is that 
the original reentrant behavior of $T_{N}$ (solid line with crosses) 
is fictitious and instead a true phase boundary is given by $T_{N}^{\rm IC}$. 
We find that these results are generic and applicable for various 
band parameters, which reproduce an electron-like Fermi surface 
that does not cross the magnetic Brillouin-zone boundary $|k_x|+|k_y|=\pi$ 
around $\delta = \delta_{\rm tet}$ [see \fig{phase2}(b)]. 
In this sense, the phase diagram of \fig{phase}(a), where AF with $\vq=\vQ$ 
is stabilized, should be regarded as a special case requiring a tuning of 
band parameters as already mentioned above. 
In fact, if we change the value of $t/J$ in \fig{phase} to $t/J=3$, 
keeping $t'/t$ and $t''/t$ unchanged, 
the Fermi surface stays almost the same, but $\delta_{\rm tet}$ shifts to be a bit larger. 
Such a small shift is sufficient to degrade the nesting condition of $\vq=\vQ$ 
around $\delta=\delta_{\rm tet}$. 
The resulting $T_{N}$ exhibits reentrant behavior, which is however preempted 
by $T_{N}^{\rm IC}$, similar to \fig{phase2}(a).    
There would also be a possibility that the reentrant behavior of $T_{N}$ 
shown in \fig{phase2}(a) could be preempted by a first order transition to 
AF with $\vq=\vQ$. However we checked that such a possibility does not occur 
by observing that the Landau free energy features a single minimum as a function of $m$.

\begin{figure} [t]
\includegraphics[width=10.0cm,clip]{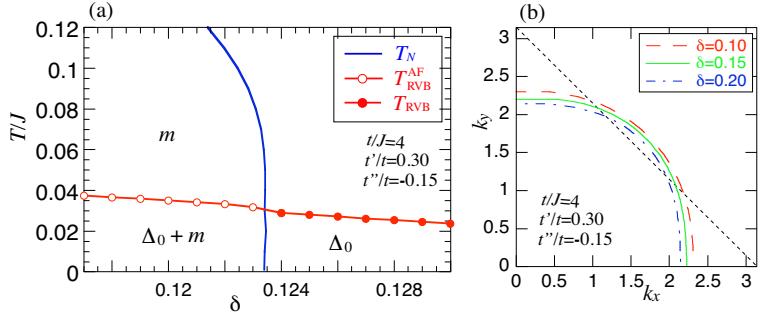}
\caption{(Color online) (a) Phase diagram in the plane of $\delta$ and $T$ 
for $t/J=4$, $t'/t=0.3$, and $t''/t=-0.15$. The notations follow those in \fig{phase}(a). 
(b) Fermi surfaces. The boundary of the magnetic Brillouin zone is denoted by 
a dotted line. 
} 
\label{phase3}
\end{figure}

If an electron-like Fermi surface crosses the magnetic Brillouin-zone boundary 
around $\delta=\delta_{\rm tet}$ as shown in \fig{phase3}(b), which may be 
applicable to electron-doped cuprates,\cite{king93,armitage02} 
$T_{N}$ tends to exhibit a straight line but still features a continuous phase transition 
[\fig{phase3}(a)]. Incommensurate magnetic order is not found to be stabilized. 
On the other hand, for band parameters leading to  
a hole-like Fermi surface such as that frequently 
used theoretically for Y- and Bi-based cuprates, AF with $\vq=\vQ$ is the most stable 
around $\delta_{N}$ and the line of $T_{N}$ exhibits reentrant behavior 
as already seen in the literature.\cite{yamase04a}

\section{Discussion and conclusion}
Our obtained phase diagrams Figs.~\ref{phase} and \ref{phase2} show good agreement with 
the experimental phase diagram.\cite{mukuda08,shimizu09a} 
AF extends over a wide doping 
region and coexists with SC at low $T$. 
Not only the value of $\delta_{N}$ but also the magnitude of $m$ 
is comparable with the experimental results.\cite{mukuda08,shimizu09,shimizu09a} 
We have invoked an electron-like Fermi surface such as that shown in 
Figs.~\ref{phase}(c) and ~\ref{phase2}(b). 
We expect such a Fermi surface is stabilized as one of Fermi surfaces 
in multilayer cuprates\cite{mori06} by virtue of strong hybridization between the layers 
with the dispersion 
$\epsilon^{z}_{\vk} \propto (\cos k_{x} -\cos k_{y})^{2}$ (Ref.~\onlinecite{andersen95}).   
The angle-resolved photoemission spectroscopy (ARPES) was performed for 
Bi$_2$Sr$_2$Ca$_2$Cu$_3$O$_{10+y}$\cite{ideta10} and 
Ba$_2$Ca$_3$Cu$_4$O$_8$F$_2$,\cite{chen10} and failed to reveal 
all Fermi surfaces.
In connection with the NMR experiments,\cite{mukuda08,shimizu09,shimizu09a}  
it is desirable to perform the ARPES for Hg- and Tl-based cuprates 
and to test a possible presence of an electron-like Fermi surface.  
It is not clear which phase diagram, \fig{phase}(a) or \fig{phase2}(a), is more appropriate to  
multilayer cuprates, since the NMR does not directly discriminate 
different ordering patterns of magnetism. Hence it is also desirable to perform 
neutron scattering measurements for the multilayer cuprates 
to reveal the wave vector of magnetic order. 

The singlet paring formation of spinons is interpreted as the pseudogap in the 
slave-boson formalism in the underdoped region. 
Since the optimal carrier density $\delta_{\rm op}$ of multilayer cuprates is 
situated above the tetracritical point $\delta_{\rm tet}$, 
$T_{\rm RVB}$ in $\delta_{\rm tet} \lesssim \delta \lesssim \delta_{\rm op}$ 
in the phase diagram 
is interpreted as the pseudogap 
temperature $T^{*}$. The previous NMR measurements\cite{julien96,kotegawa01} 
indeed observed the pseudogap behavior of $(T_{1}T)^{-1}$ in such a doping region. 
Since $J$ is around 100-150 meV,  
the obtained value of $T_{\rm RVB}$ in Figs.~\ref{phase} and \ref{phase2} 
is small compared with the experimental observation.\cite{julien96,kotegawa01}  
This discrepancy should be explored by including explicitly the multilayer degree of 
freedom in the present analysis. 

On the other hand, 
for $\delta < \delta_{\rm tet}$,  
U(1) gauge fluctuations emerging in the slave-boson formalism 
are expected to be strongly suppressed in the antiferromagnetic state and thus 
spinons and holons tend to confine there.\cite{kim99}  
Therefore we expect that the pseudogap in the antiferromagnetic state is 
substantially diminished for $\delta < \delta_{\rm tet}$ and instead 
the coexistence of AF and SC is realized unless 
the tendency of carrier localization appears at low temperatures. 
Around the tetracritical point it is interesting to clarify both theoretically 
and experimentally how the onset temperature of the pseudogap 
changes to that of the phase transition to the coexistence with decreasing $\delta$. 

A crucial difference between single-layer and multilayer 
cuprates lies in the difference of antiferromagnetic fluctuations.  
AF is realized through breaking of continuous symmetry, which thus 
does not occur at a finite $T$ in a pure 2D system.\cite{mermin66} 
Hence the presence of AF in layered materials is interpreted as coming from 
weak three dimensionality, which is always present in real systems and 
in general suppresses fluctuations. 
In single-layer cuprates, because of the intrinsic low dimensionality 
due to a tiny coupling between CuO$_2$ layers along the $c$ axis, 
antiferromagnetic fluctuations are expected to be so strong that the magnetism is realized 
only close to the Mott insulator. 
Moreover some extrinsic effect such as randomness may easily hinder 
long-range antiferromagnetic order. The standard slave-boson formalism 
of the $t$-$J$ model (without AF) was proposed for such 
cuprate superconductors.\cite{lee06}  
On the other hand, for multilayer cuprates, 
many layers are already present within a unit cell, 
yielding relatively strong three dimensionality compared with 
single-layer systems. 
In addition, each CuO$_{2}$ plane is perfectly flat and free from 
disorder. These can be the main reasons why 
the present slave-boson mean-field analysis with AF 
captures essential features observed in multilayer cuprates.\cite{mukuda08,shimizu09,shimizu09a}  
A natural consequence is that AF would extend over a wider doping region by 
increasing the number of CuO$_{2}$ planes in a unit cell 
unless the value of $t/J$ varies significantly. This tendency is actually reported 
in Ref.~\onlinecite{shimizu09}. 

Multilayer cuprate superconductors achieve 
much higher $T_{c}$ than single-layer systems. Hence 
the understanding of multilayer cuprates is 
crucially important to elucidate the mechanism of high-temperature SC. 
We have argued that weak three dimensionality coming from 
a multilayer structure is sufficient to stabilize AF and that 
multilayer cuprates can be systems described by RVB and AF in the $t$-$J$ model, 
suggesting the importance of the local antiferromagnetic coupling $J$.

\begin{acknowledgments}
We are grateful to H. Mukuda for useful discussions. 
H.Y. thanks O. K. Andersen, M. Fujita, V. Hinkov, A. A. Katanin, W. Metzner, and R. Zeyher 
for valuable discussions. 
\end{acknowledgments}


\bibliography{main.bib}

\end{document}